\documentclass[10pt]{article}

\usepackage{multicol}
\usepackage{subfigure}
\usepackage{mathtools}
\usepackage{graphicx}
\usepackage{amsmath}
\usepackage{mathrsfs}
\usepackage[top=1in,bottom=1in,left=1in,right=1in]{geometry}
\usepackage{pdflscape}
\usepackage{times}
\usepackage{bm}
\usepackage{color}
\usepackage{caption}
\usepackage{amsmath}
\usepackage{amssymb}
\usepackage{listings}
\usepackage{textcomp}
\usepackage{xcolor}
\usepackage{algorithm2e}
\usepackage{float}
\usepackage{algorithmicx}
\usepackage{algpseudocode}
\usepackage{hyperref}
\usepackage{sectsty}
\sectionfont{\LARGE}
\subsectionfont{\LARGE}
\subsubsectionfont{\Large}

\hypersetup{hidelinks,
	colorlinks=true,
	allcolors=black,
	pdfstartview=Fit,
	breaklinks=true}

\begin{document}
\pagenumbering{arabic}

\thispagestyle{plain}
\newpage
\setcounter{page}{1}

\begin{center}



\huge \textbf{Structured multi-factor risk models}

\begin{center}
\includegraphics[width=0.8\textwidth]{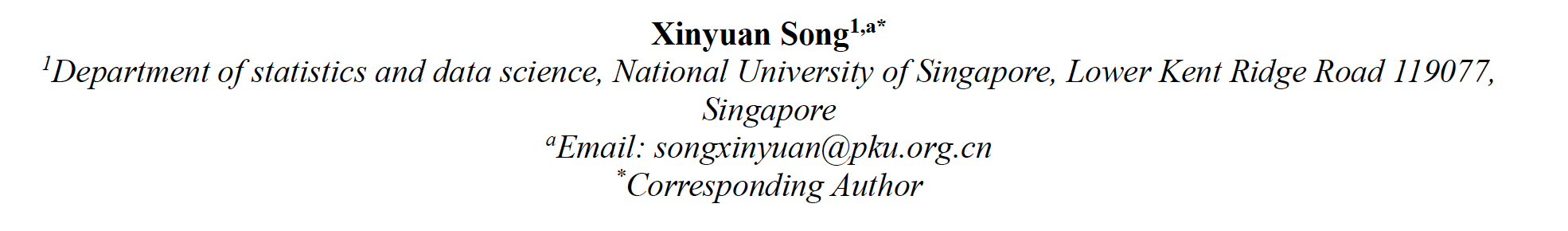}  
\end{center}


\end{center}

\begin{center}
    \huge Abstract
\end{center}

\Large
Income and risk coexist, yet investors are often so focused on chasing high returns that they overlook the potential risks that can lead to high losses. Therefore, risk forecasting and risk control is the cornerstone of investment. To address the challenge, we construct a multi-factor risk model on the basis of the classical multi-factor modelling framework. \\

For the common factors, inspired by Barra Model's factor classification, we select 10 style factors, 29 industry factors and 1 country factor. To ensure the model is solid, we adjust the outliers and missing values of factor exposure data, normalize and finally orthogonalize them, before computing factor returns and making further analysis. 
Factor return covariance matrix and idiosyncratic return variance matrix are essential tools to express stock returns in the multi-factor risk model. Firstly, we calculate the factor return covariance matrix with EWMA. To tackle the time-series autocorrelation of factor returns, we apply Newey-West adjustment. Then we estimate the idiosyncratic return variance matrix in a similar way and make Newey-West adjustment again to solve the time-series autocorrelation problem. Since the return of a single share is sensitive to missing values and outliers, we introduce structural adjustment to improve the matrix.
Eventually, we obtain the return covariance matrix among stocks and compute the risk of investment portfolio based on it.\\

Furthermore, we search for optimal portfolio with respect to minimizing risk or maximizing risk-adjusted return with our model. They provide good Sharpe ratio and information ratio for considering both absolute risk and active risk. Hence, the multi-factor risk model is efficient.

\newpage
\tableofcontents

\newpage
\section{Introduction}

In the investment process, often higher profits come with greater risk. And investors are always striving for the highest returns with the lowest risk. In this case, having models that do an accurate job of predicting risk would greatly enhance the return to investors. Accurate risk prediction is the cornerstone of risk control.\\
The Markowitz mean-variance model is a widely used model in the field of risk. It provides the basis for risk prediction, but the model has some limitations in practice. Firstly, directly calculating covariance matrix between stock returns is time-consuming. Secondly, because of estimation error, the length of the sample period must be larger than the number of stocks to get good estimation result. Thirdly, the Markowitz mean-variance model is very sensitive to input changes and not robust enough.\\
Because of the limitations above, structured multi-factor risk models have become a powerful tool for analysing portfolio risk. By converting covariance and variance estimation of stocks with high dimensions into covariance and variance estimation of factors with significantly lower dimensions, structured multi-factor risk models not only reduce the computational effort but also improve the accuracy and robustness of risk prediction.\\
This paper mainly focus on the construction methods of multi-factor risk models and their application in portfolio optimization. The first section describes the selection of common factors and the calculation of factor returns for risk models. The second section is about estimation and adjustment methods for the factor return covariance matrix and the idiosyncratic return variance matrix. Finally, the last section will assess the performance of the risk model when used in portfolio optimization.

\section{Structured Multi-factor risk models}
A structured multi-factor risk model represents stock return as a combination of returns from a set of common factors and its idiosyncratic return related only to that stock:
\begin{equation}
    r_n=\sum_{k=1}^{k}X_{n k}f_k+u_n
\end{equation}
where $r_n$ is the return of $n^{th}$ stock, $X_{nk}$ is the factor exposure of $n^{th}$ stock on $k^{th}$ factor. $f_k$ is the $k^{th}$ factor’s return and $u_n$ is the $n^{th}$ stock’s idiosyncratic return.\\ 
Stock return risk is a combination of common factor return risk and idiosyncratic return risk. The idiosyncratic return risk represents the volatility of returns that cannot be explained by the common factors, such as changes in share prices driven by certain unexpected events, and unknown alpha factors. Under this condition, common factor returns and idiosyncratic returns are not correlated with each other. \\
The multi-factor risk model converts the estimation of the stock return covariance matrix into an estimate of the factor covariance matrix and the idiosyncratic return variance matrix(Richard \& Ronald,1994).:
\begin{equation}
    V=XFX^T+\Delta
\end{equation}

\centerline{$V$: the covariance matrix between stock returns.}
\centerline{$X$: the factor exposure matrix for stocks. }
\centerline{$F$: the covariance matrix between factor returns}
\centerline{$\Delta$: the equity-idiosyncratic return variance matrix}
In this article, we build a multi-factor risk model based on Barra multi-factor risk model. We choose appropriate and valid common factors, regresses them to calculate and estimate the covariance matrices. In addition, we make some necessary adjustments to the two matrices to improve their estimation accuracy, and thus to improve our risk model. Finally, we tested our risk model in portfolio optimization with different constraints and evaluated performance of the portfolios we obtain.\\

\subsection{Factor choosing}
 The selection of appropriate and effective co-factors is the basis for building an effective risk model. Firstly, the factors must be able to explain stock returns consistently and effectively. Factors with little explanatory power provide very limited incremental information. Also, the random noise they carry will directly affect the stability of risk prediction. In addition, we wish to construct the model with the least number of factors possible to improve model efficiency in computation.
 
 In our paper, based on Barra Model's factor classification, we choose 10 style factors($f_S$), 29 industry factors($f_i$) and 1 country factor($f_C$). 
 
\subsubsection{Style Factor}
\begin{figure}
    \centering
     \includegraphics[width=1\textwidth]{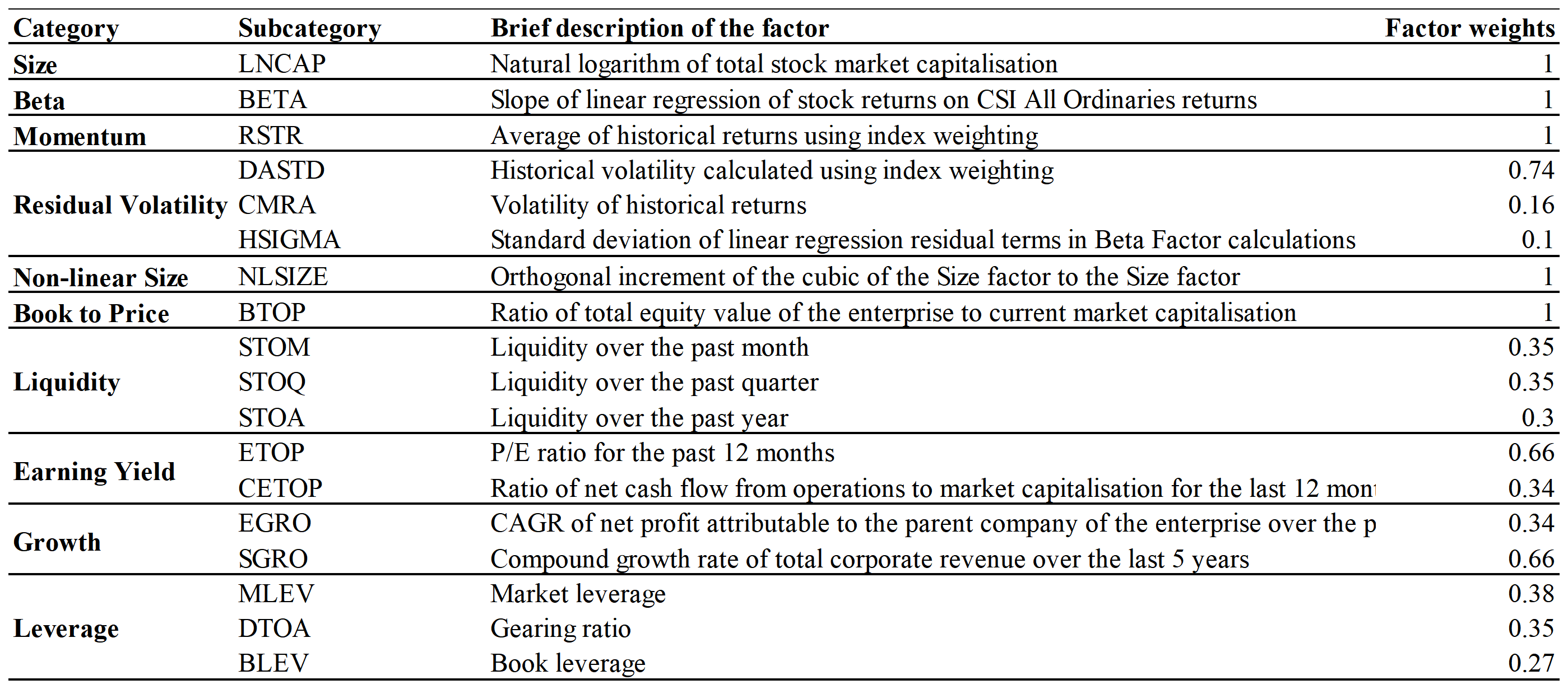}
    \caption{Style factors}
    \label{fig:fs}
\end{figure}

The Style Factor (figure \ref{fig:fs}) consists of 10 broad fundamental factors, namely Size, Beta, Momentum, Residual Volatility, Non-linear Size, Book to Price, Liquidity, Earning Yield, Growth and Leverage. The factors are represented as a weighted combination of several sub-categories, thus solving the possible problem of col-linearity of the sub-categories while enriching the information as much as possible.

\subsubsection{Industry Factor}
The industry factors provide a rich source of incremental information to the model in addition to the style factors. Based on the 29 CITIC primary sectors in the A-share market, the model includes 29 industry factors. They are coal, transportation, real estate, power and utilities, machinery, power equipment, non-ferrous metals, basic chemicals, trade and retail, construction, light manufacturing, general, pharmaceuticals, textiles and garments, food and beverage, household appliances, automobiles, electronic components, building materials, food and beverage, petroleum and petrochemicals, defence and military, agriculture,  steel, communications, computers, non-banking finance, media and banking. 

\subsubsection{Country Factor}
Traditional multi-factor models typically include style factors and industry factors, but here we refer to Barra's approach and explicitly include country factors in the multi-factor risk model:
\begin{equation}
    r_n=f_c+\sum_{i} X_{ni}f_i+\sum_{s}X_{ns} f_s+u_n,
\end{equation}
\centerline{$f_c$ : the country factor return}
\centerline{$X_{ni}$ : the factor exposure (0 or 1) of the $n^{th}$ stock on the $i^{th}$ industry factor.}
\centerline{$f_{i}$ : the return on the $i^{th}$ industry factor. }
\centerline{$X_{ns}$ : the factor exposure of the $n^{th}$ stock on the $s^{th}$ style factor}
\centerline{ $f_s$: the return on the $s^{th}$ style factor.}

To eliminate the collinearity between the country and industry factors, the model needs to add additional constraints to the industry factors:
\begin{equation}
    \sum^{29}_{i=1}w_i f_i=0
\end{equation}
$w_i$ is the weight of $i^{th}$ industry's total circulating market capitalization. 
Barra USE4(Jose, D.J. \& Jun, 2011) compares the correlation coefficients of the industry factors in the models with and without the country factor. Result shows that correlation means of industry factors for the two models are similar, but the model with country factor is more sensitive to market movement based on time series.

\subsection{factor return}
\subsubsection{Factor exposure data processing}
Data is the foundation of a model. Therefore, before obtaining factor return, we make the following adjustment to the exposure of style factors. As Figure 2 shows, there are 4 steps to process the factor exposure.
\begin{figure}[H]
    \centering
     \includegraphics[width=1\textwidth]{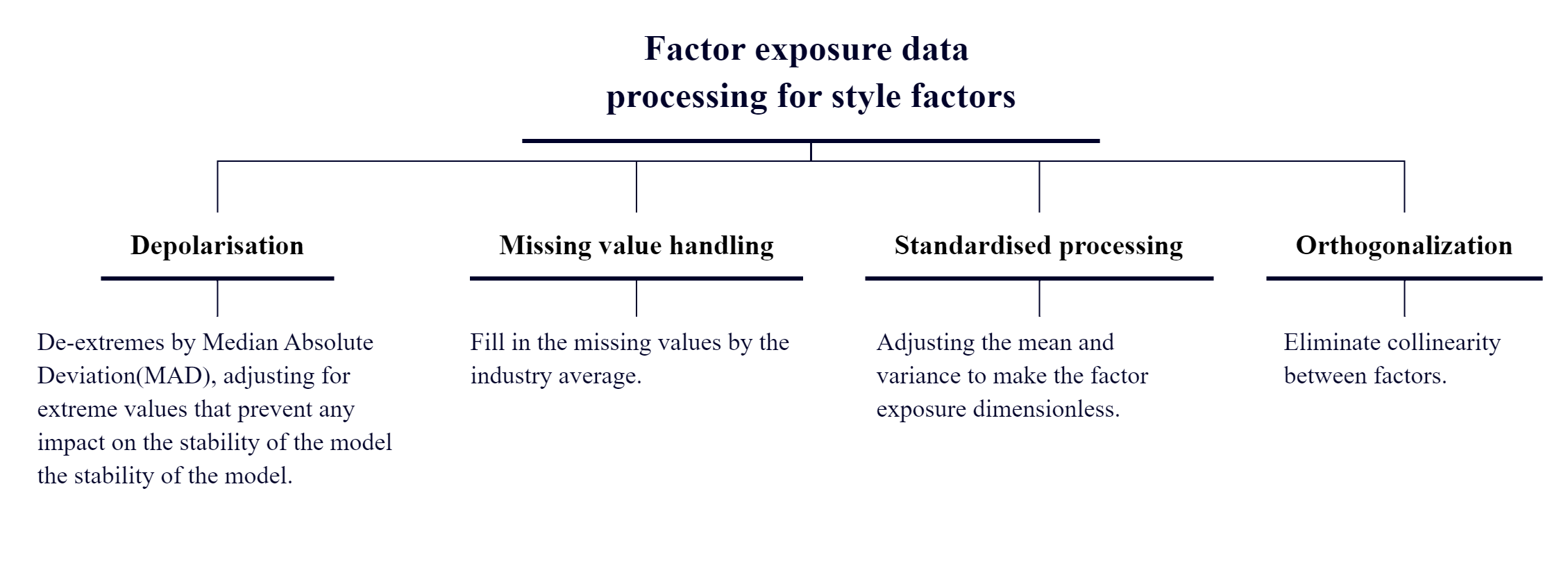}
    \caption{Factor exposure processing step}
    \label{fig:Fig2}
\end{figure}

\begin{itemize}
    \item Step 1: Depolarisation
\end{itemize}

Some extraordinarily skewed data can affect the accuracy of our model, where we use Median Absolute Deviation(MAD). Denote $D_k$ as the exposure series of the $n^{th}$ factor on all individual stocks at a time cross-section. $D_M$ is the median of $D_k$. $D_{MAD}$ is the median of series $|D_M-D_K|$ and $\widetilde D_K$ is the exposure after depolarisation. 
\begin{equation}
    \widetilde D_k=
\left \{
    \begin{array}{lr}
    D_M+3*D_{MAD}     & if D_i>D_M+3*D_{MAD} \\
    D_M-3*D_{MAD}     & if D_i<D_M-3*D_{MAD} \\
    D_k &else
    \end{array} 
\right.
\end{equation}

\begin{itemize}
    \item Step 2:Missing value handling
\end{itemize}

Missing values can appear in different time cross-sections, different stocks and for different factors. We can directly move the factors with missing values out of the model, but it will damage the model effectiveness and precision since the useful information of the dropped factor is not include in the model. Filling the missing values with industry average can assure the number of factors is consistent with respect to different time cross-sections and help us get a more effective and stable model.

\begin{itemize}
    \item Step 3: Normalization
\end{itemize}

Standardization of factor exposure data for style factors:
\begin{center}
    $d_{nk}=\frac{d_{nk}^{raw}-u_k}{\sigma_k}$
\end{center}
Here, $d_{nk}^{raw}$ is the factor exposure of the $k^{th}$ factor on the $n^{th}$ stock after we processing step 2. \\
The standardized factor exposures follow a roughly $N(0,1)$ standard normal distribution, where the magnitudes of the factors are eliminated and comparability between factors is achieved. At the same time, the model makes clearer economic sense. The returns on the standardized factors, calculated by linear regression, actually reflect the returns on the pure factor portfolios of the factors.

\begin{itemize}
    \item Step 4: Orthogonalization process
\end{itemize}

Collinearity between style factors is addressed through orthogonalization. \\
Orthogonalization is achieved by linearly regressing the factor exposure of the factor to be treated on the factor exposure of the selected factor at each cross section, using weighted least squares (WLS) method with market cap as weights. Then, regard the residuals of the regression as the factor exposure of the factor after orthogonalization.\\
It is important to note that the orthogonalization process should not be overused. The orthogonalization of a weakly correlated factor can lead to a technical analysis result but lose its economic meaning. Therefore, only two factors are orthogonalized here: the Residual Volatility factor is orthogonalized to the Size and Beta factors, and the Liquidity factor is orthogonalized to the Size factor. The orthogonalized factors no longer satisfy the standard normal distribution and need to be normalized again.

\subsubsection{Calculation of factor returns}
We need to solve the following optimization problem to obtain factor returns.
\begin{align}
 Min\ Q & =\sum^N_{n=1} w_n(r_n-f_c-\sum_i X_{ni}f_i-\sum_s X_{ns}f_s)^2 \\
s.t. & \sum^N_{n=1} \sum_i w_n X_{ni} f_i=0 \notag
\end{align}

We can firstly consider the model without country factor:
\begin{align}
 Min\ Q =\sum^N_{n=1} w_n(r_n-\sum_i X_{ni} \widetilde f_l-\sum_s X_{ns}f_s)^2 
\end{align}
This optimization problem can be solved using weighted least squares, and we can get style factor returns and industry factor returns without considering country factor. Then we set the country factor return to be equal to the market cap weighted average of the industry factor returns, and subtract the country factor return from industry factor returns. In this way, we obtain the country factor returns and industry factor returns that satisfy the restricts in the full model.
\begin{align*}
f_c &=\sum^N_{n=1} \sum_i w_n X_{ni} \widetilde f_l\\
f_i &=\widetilde f_l - f_c
\end{align*}

\section{Estimation of the risk matrix}
The multi-factor risk model explain stock return covariance matrix using factor return covariance matrix and idiosyncratic return variance matrix. And we are going to compute and adjust these two matrices in the following section.\\
To evaluate the precision of risk prediction, we introduce the following bias statistic. Define the out-of-sample standardized return for asset $k$ as follows:\\
\begin{align}
    b_{k,t}=\frac{r_{k,t\sim t+\delta}}{\sigma_{k,t}},
\end{align}
where $r_{k,t\sim t+\delta}$ is the factor return from the current $t^{th}$ cross-section to the future $t^{th}+\delta$ cross-section on asset $k$. And $\sigma_{k,t}$ is the model's predicted value of return volatility for asset $k$ in the cross-sectional $t^{th}$ to $t^{th}+\delta$ interval at $t$.\\

The bias statistic is defined as the standard deviation of standardized return for a certain prediction period, denoted as $B_k$:
\begin{align}
    B_k=\sqrt{\frac{1}{T-1}\sum^T_{t=1}(b_{k,t}-\overline{b_k})}
\end{align}
and $\overline{b_k}$ is the average of $b_{k,t}$ in period $T$.
Bias statistic represents the ratio of actual risk to predicted risk for the asset. If the predicted risk is the same as actual risk, $B_k=1$. If risk is underestimated, $B_k>1$. If risk is overestimated, $B_k<1$. In reality, because of sampling error, the bias statistic itself is a random variable. Even if the risk prediction is perfect, the bias statistic will still fluctuate around 1.

\subsection{Factor return covariance matrix}
To obtain a precise factor return covariance matrix from historical returns, the number of time cross-sections need to be larger than number of factors. If we use monthly data, we’ll need records of at least the past 3 years to obtain solid estimate for a model with 40 style factors. A long historical horizon will make the model involve too much past information that is not relevant to the current market. And the low frequency of monthly data means they contain much less information than daily data, making the model hard to capture changes in the market in time. As for daily data, they contain more fresh information and directly reflect market changes. Considering the aspects above, we choose to estimate factor return covariance matrix and idiosyncratic return variance matrix using daily data.
In addition, we assign larger weights to the more recent records to further avoid too much past information in the model. And we use the following Exponentially Weighted Moving Average Method to obtain factor return covariance matrix $F^{Raw}$:
\begin{align}
F_{a,b}^{Raw}=cov\left( f_a,f_b \right) _t=\sum_{s=0}^h{\lambda _{t-s}\left( f_{a,t-s}-\bar{f}_b \right)}/\sum_{s=0}^h{\lambda _{t-s}}
\end{align}
$$
\lambda _{t-s}=0.5^{s/\tau}
$$
$cov\left( f_a,f_b \right) _t$: the covariance between factor $a$ and factor $b$ in period $t$. \\
$\lambda $: the index decay weights. \\
$f_{a,t-s},f_{b,t-s}$: returns of factor $a$ and factor $b$ at the $t-s$ cross-section. \\
$\overline{f}_a,\overline{f_b}$: weighted mean of the returns of factor $a$ and factor $b$ in the interval from cross-section $t-h$  to cross-section $t$.\\
Here we set the length of the time horizon $h=252$, half life of the weight $\tau=90$.

\subsubsection{Newey-West adjustment}
The covariance matrix above $F^{Raw}$ is obtained under the assumption that factor returns have no auto correlation. In fact, factor returns do have auto correlation so we need to include this feature in our model.
The Moving Average model is a good choice to capture auto correlation in factor returns. Assume factor returns have auto correlation of order $D$, then the factor return can be represented as follows:
\begin{align}
f\left( t \right) =\mu +\sum_{i=0}^D{\theta _i\varepsilon _{t-i}}
\end{align}

The white noise $\varepsilon _{t-i}$ represents all the information that influences factor return at time $t-i$, and this kind of influence will exist $D$ periods. For the factor return matrix with auto correlation of order $D$, a simple estimator is as follows:
\begin{align}
\varOmega =F^{Raw}+\sum_{d=1}^D{\left( \hat{\varOmega}_d+\hat{\varOmega}_{d}^{\prime} \right)}
\end{align}
$$
\hat{\varOmega}_d=\sum_{t=1}^{T-d}{\lambda ^{T-d-t}f_tf_{t+d}^{\prime}}/\sum_{t=1}^{T-d}{\lambda ^{T-d-t}}
$$
In this method, we use the 1~$d$ order autocovariance matrix to adjust the covariance matrix without time-series autocorrelation, $F^{Raw}$. But the updated result may not be positive semi-definite. To solve this problem, we use Newey-West adjustment:
\begin{equation*}
\hat{\Omega} = F^{Raw}+\sum_{d=1}^{D}w(d,D)\cdot(\hat{\Omega}_d+\hat{\Omega}^{'}_d) 
\end{equation*}
\begin{equation*}
w(d,D) = 1-\frac{d}{D+1}
\end{equation*}
\begin{center}
$w(d,D)$: the Bartlett weight, which decrease with increasing $d$
\end{center}
The covariance matrix after Newey-West adjustment is a consistent estimator and is positive semi-definite.
The factor return covariance matrix here is obtained based on daily data and represents the risk of daily return. When considering risk of monthly return, we need to convert the daily return covariance matrix to the monthly return covariance matrix:
\begin{equation*}
F^{NW} = 21\cdot\hat{\Omega}=21\cdot[F^{Raw}+\sum_{d=1}^{D}(1-\frac{d}{D+1})\cdot(\hat{\Omega}_d+\hat{\Omega}^{'}_d)]
\end{equation*}
Here we set the lag order $D=2$.

\begin{figure}[H]
    \centering
    \includegraphics{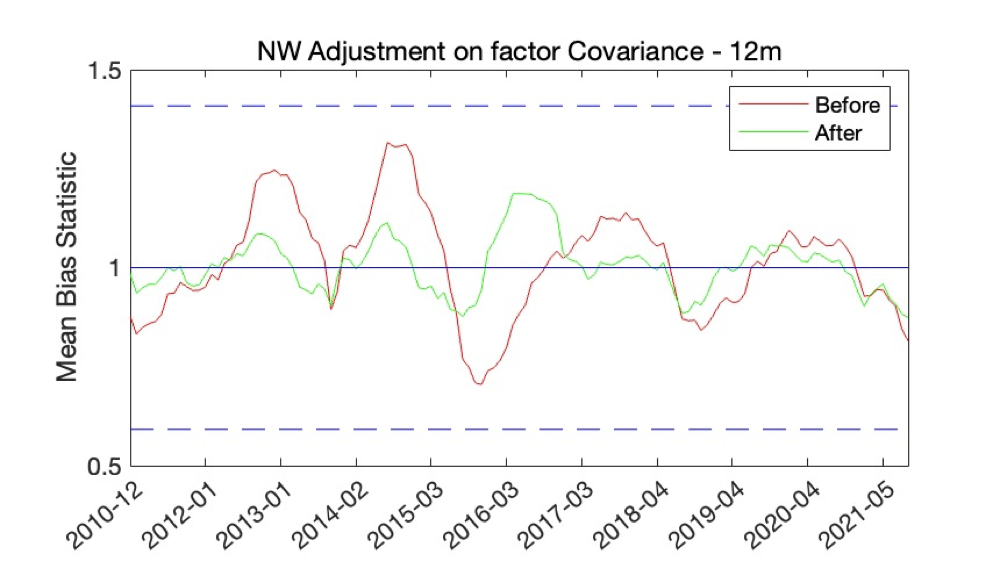}
    \caption{Newey-West adjustment on factor return covariance matrix}
    \label{fig:NW on factor}
\end{figure}

Figure \ref{fig:NW on factor} shows the improvement of Newey-West adjustment on factor return covariance matrix. The green line shows that after the adjustment, the bias statistic becomes closer to 1, indicating improvement in volatility prediction accuracy.

\subsection{Idiosyncratic return variance matrix}
The idiosyncratic return variance matrix$(\Delta)$ is another major component of a multi-factor risk model. Idiosyncratic returns of different stocks are independent and uncorrelated with each other, so idiosyncratic return variance matrix is a diagonal matrix. For consistency, we use the same EWMA method to calculate the idiosyncratic return variance matrix based on daily return data:
\begin{align}
    \hat{\Omega_0} &=var(u_n)_t = \frac{\sum^h_{s=0}  \lambda_{t-s}(u_{n,t-s}-\bar{u_n})^2 }{ \sum_{s=0}^h \lambda_{t-s}}\\
    \lambda_{t-s} &=0.5^{\frac{s}{\tau}} 
\end{align}

$u_{n,t-s}$: the idiosyncratic return of the nth stock on the $t-s$ cross section,\\
$u_n$: The weighted mean of the idiosyncratic return of stock n in the interval from section $t-h$ to section $t$,\\
Same as the factor return covariance matrix, we set the time window $h = 252$ and the weight half-life $\tau = 90$.\\

\subsubsection{Newey-West adjustment}
Similarly, due to the time-series autocorrelation, it is necessary to conduct the Newey-West adjustment first to modify the estimated idiosyncratic return variance matrix. And then convert to the monthly one that characterizes monthly risk:
\begin{align}
    (\sigma^{NW})^2=21 \cdot [\hat{\Omega_0}+\sum^D_{d=1}(1-\frac{d}{D+1}) \cdot(\hat{\Omega_d}+\hat{\Omega'_d})]
\end{align}
Here the lag is $D = 5$.

\begin{figure}[H]
    \centering
    \includegraphics[height=0.8\textwidth]{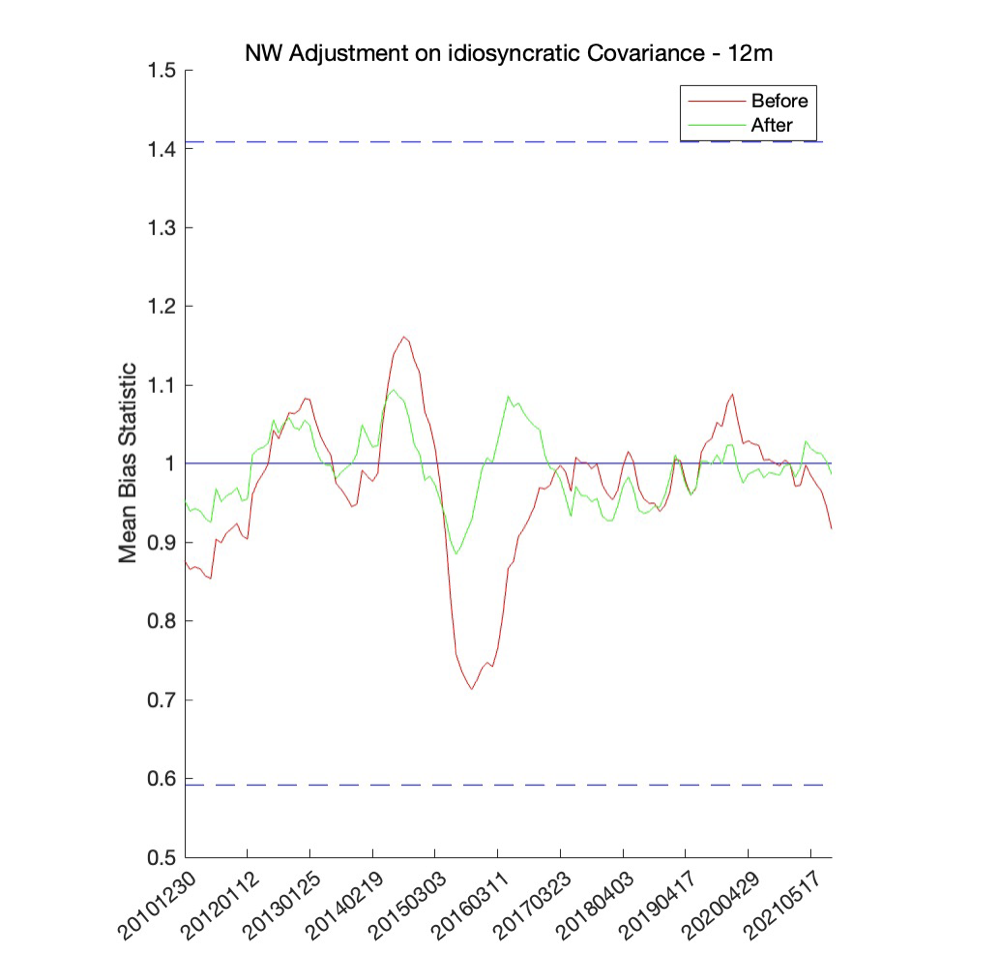}
    \caption{Newey-West adjustment on idiosyncratic covariance matrix}
    \label{fig:NW on idio}
\end{figure}
Figure \ref{fig:NW on idio} shows the improvement of Newey-West adjustment on idiosyncratic return covariance matrix. The improvement in volatility prediction accuracy is also quite noticeable, with the bias statistic becoming closer to 1.
\subsubsection{Structural adjustment}

At one particular time cross-section, the factors' returns are calculated by linear regression of thousands of stocks, hence they are not easily affected by missing values or outliers of one single stock. However, idiosyncratic returns are calculated based on a single stock, and they can be easily affected by missing values or outliers. In practice, the idiosyncratic returns of newly listed stocks and stocks that have been suspended for a long time may be missing. In addition, the idiosyncratic return of the stock may generate outliers near important company announcement. These missing data and outliers may cause the idiosyncratic return variance matrix to be non-stationary, and reduce the stability of the model.
Therefore, for the idiosyncratic return variance matrix, we need to set up an additional structural adjustment to correct the impact of  missing values and outliers on the risk matrix.

Assumed that stocks with the same factor exposure may have the same idiosyncratic risk, the specific method of structural adjustment is as follows: For an idiosyncratic return series within a time horizon ($h$ = 252), define the coordination parameter of the $n^{th}$ stock $\gamma_n \in [0,1]$:
\begin{align}
    \gamma_n &=\{min[1,max(0,V_n)]\} \cdot \{min[1,exp(1-Z_n)]\}\\
    V_n &=\frac{h_n-60}{120}\notag \\
    Z_n &=|\frac{\sigma_n-\widetilde\sigma_n}{\widetilde\sigma_n}| \notag
\end{align}
$V_n$: shows the degree of missing data, the larger the $V_n$, the smaller the degree of missing data.\\
$h_n$: the effective data in the sample, if $h_n \geq 180$, it is considered that there is no obvious missing data.\\
$Z_n$: the degree of heavy tail, the greater the $Z_n$ is greater than 1, the greater the degree of heavy tail.\\ 
$\sigma_n, \widetilde\sigma_n$: The equal weighted standard deviation of the sample, the robust estimated standard deviation of the sample.\\
$Q_{3,n}, Q_{1,n}$: three-quartile of the sample, the quartile of the sample.\\
If the $n^{th}$ stock has a heavy tail caused by obvious missing data or outliers in this time horizon, then $\gamma_n < 1$. Otherwise, $\gamma_n = 1$.\\

For all stock data with $\gamma_n = 1$, we can do linear regression on the log of the volatility of stock-idiosyncratic returns $\sigma_n^{TS}$ to the factor exposure of all factors. The regression adopts the weighted least squares $(WLS)$ method weighted by market cap to obtain the contribution value $b_k$ of each factor to the idiosyncratic volatility:
\begin{align}
    ln(\sigma^{TS}_n)=\sum_k X_{nk} \cdot b_k+\epsilon_n
\end{align}

Then, the predicted value of structural idiosyncratic volatility for the $n^{th}$ stock $\sigma_n^{STR}$ is:
\begin{align}
    \sigma_n^{STR}=E_0 \cdot exp(\sum_k X_{nk} \cdot b_k)
\end{align}
where $E_0$ is the adjustment coefficient to eliminate the influence of $\epsilon_n$. Here we set $E_0$ = 1.05.\\
For all individual stocks, we use $\gamma_n$ as the weight to calculate the new idiosyncratic fluctuations after structure adjusted:
\begin{align}
    \hat{\sigma_n}=\gamma_n \cdot \sigma_n^{TS}+(1-\gamma_n)\cdot \sigma_n^{STR}
\end{align}

The structural adjustment is only for stocks with obvious missing values or outliers, that is, $\gamma < 1$. For stocks with good data quality ($\gamma = 1$), the idiosyncratic volatility before and after structural adjustment remains unchanged.

\begin{figure}[htbp]
\centering
\begin{minipage}[t]{0.48\textwidth}
\centering
\includegraphics[width=7.5cm]{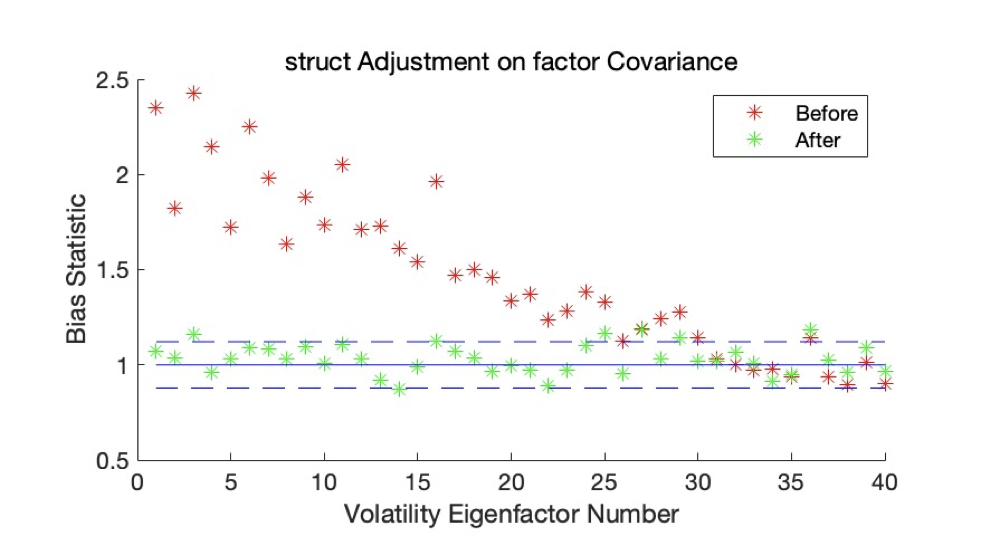}
\caption{Structure adjustment on factor return covariance matrix}
\label{fig:strf}
\end{minipage}
\begin{minipage}[t]{0.48\textwidth}
\centering
\includegraphics[width=7.5cm]{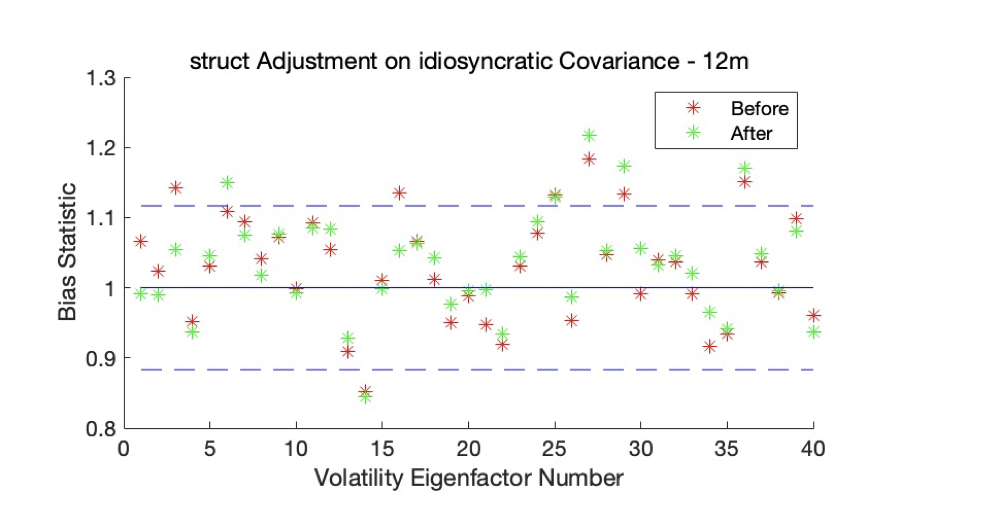}
\caption{Structure adjustment on idiosyncratic covariance matrix}
\label{fig:stri}
\end{minipage}
\end{figure}

Figure \ref{fig:strf} and figure \ref{fig:stri} shows separately the improvement of the bias statistics from the two matrices after the structural adjustment. And we can see that the volatility prediction accuracy improved a lot after both adjustments, with the bias statistic becoming closer to 1. For the factor return covariance matrix adjustment, the forecast error improved considerably. For the idiosyncratic variance matrix adjustment, the improvement was modest and there was no significant difference after the adjustment.

\begin{figure}
    \centering
    \includegraphics[width=0.8\textwidth,height=0.6\textwidth]{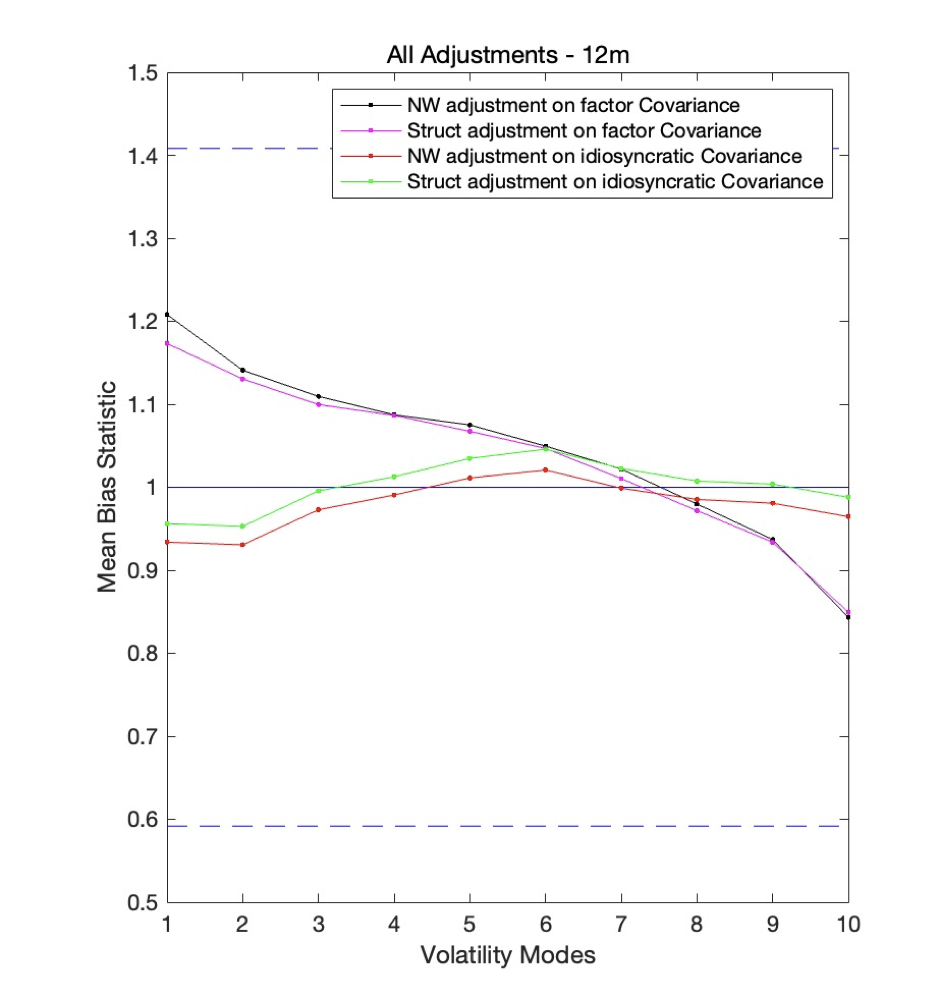}
    \caption{All adjustment comparison}
    \label{fig:allad}
\end{figure}
And here we divide all stocks into 10 different groups according to their forecasted volatility, and calculate the mean bias statitic for each group using models with different adjustment. We plot all our adjustment results in the same graph for comparison (figure \ref{fig:allad}). Each adjustment offers a different degree of enhancement. It can be seen that the Newest-West adjustment provides a better lift for both matrices, while the structured adjustment is significantly more effective for the idiosyncratic return variance matrix.

\newpage
\section{Application of multi-factor risk model in portfolio optimization}
There are two main applications of multi-factor risk model, to help construct optimal portfolio in optimization and to analyze attribution of portfolio return to its risk. We only focus on the former one, and try to obtain optimal portfolio with our risk model.
Firstly, we simply test our risk model by forecasting the volatility of CSI 500 index, and by constructing a minimal risk portfolio based on the CSI 500 index components. The results are as follows.

\begin{figure}[H]
    \centering
     \includegraphics[width=0.8\textwidth]{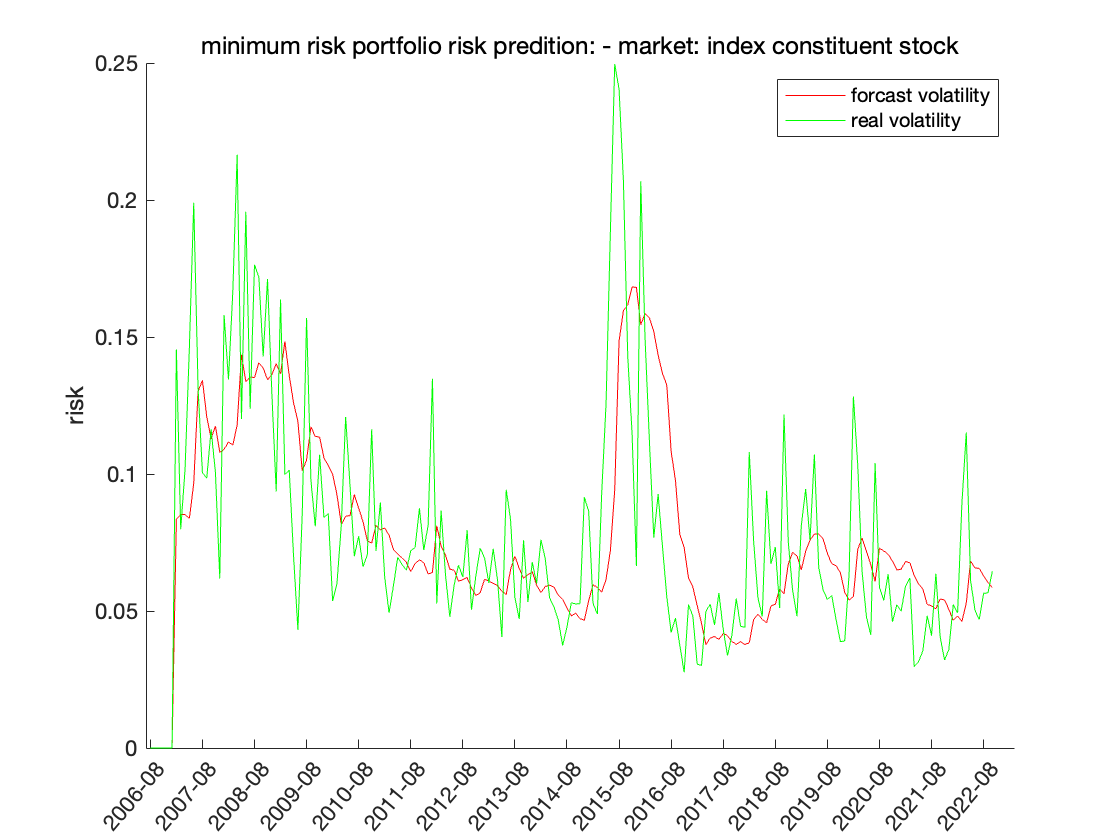}
    \caption{CSI 500 index volatility forecast}
    \label{fig:Fig8}
\end{figure}

\begin{figure}[H]
    \centering
     \includegraphics[width=0.8\textwidth]{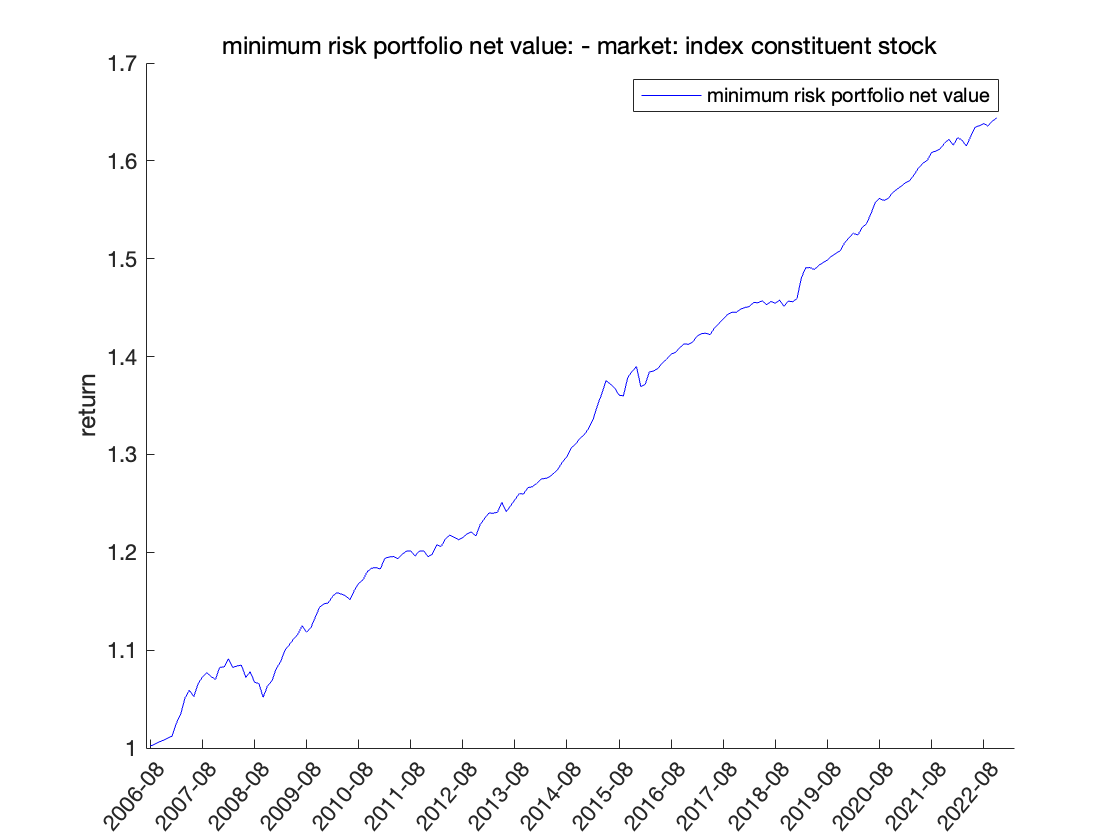}
    \caption{CSI 500 minimum risk portfolio's daily return - stockmarket: index constituent stocks}
    \label{fig:Fig9}
\end{figure}

We can see that the forecasted volatility of CSI 500 index is close to the real volatility, and the portfolio we construct is profitable. This roughly indicates that our risk model is effective in volatility forecast and is effective in the process of portfolio optimization.

\subsection{Minimum risk portfolio}
In multi-factor risk model, risk of portfolio $P$ with weight $w$ is expressed as below.
~\\
\begin{equation}
    \sigma^2_P(w)=w^TVw=w^T(XFX^T+\Delta)w
\end{equation}
\centerline{X: factor exposure matrix for all stocks}
\centerline{F: factor return covariance matrix}
\centerline{$\Delta$: idiosyncratic return variance matrix}
~\\
Hence, the object function of minimizing risk is:
\begin{align}
    Min_w\ w^T(XFX^T+\Delta)w
\end{align}
We can select the forms of weights $w$ based on the optimization object. If the object is to minimize the absolute risk of portfolio, $w$ should be the original portfolio weight. If we want to maximize active risk, $w$ should be the difference of weight compared with the benchmark.
Besides, additional constraints can be incorporated into the optimization, such as non-negative portfolio weights and only to select stocks from certain stock pool.

\subsubsection{Optimize portfolio weight to minimize absolute risk}
Here $w$ is the portfolio weight and we want to construct portfolio with minimum absolute risk:
\begin{equation}
\begin{split}
    Min_w\ & w^T(XFX^T+\Delta)w\\
    s.t. &\ \forall n\ w_n\ge0\\
    \Sigma_n &w_n=1
\end{split}
\end{equation}
Here we construct minimum absolute risk portfolios based on the stock pool of CSI 500 index. Our portfolio has monthly turnover, and the position is adjusted at the beginning of every month. With the backtesting period from 2011 to 2020, the graph below shows portfolio's net value, cumulative net value of excess return, drawdown and other assessment indicators.

\begin{figure}[H]
    \centering
     \includegraphics[width=0.8\textwidth]{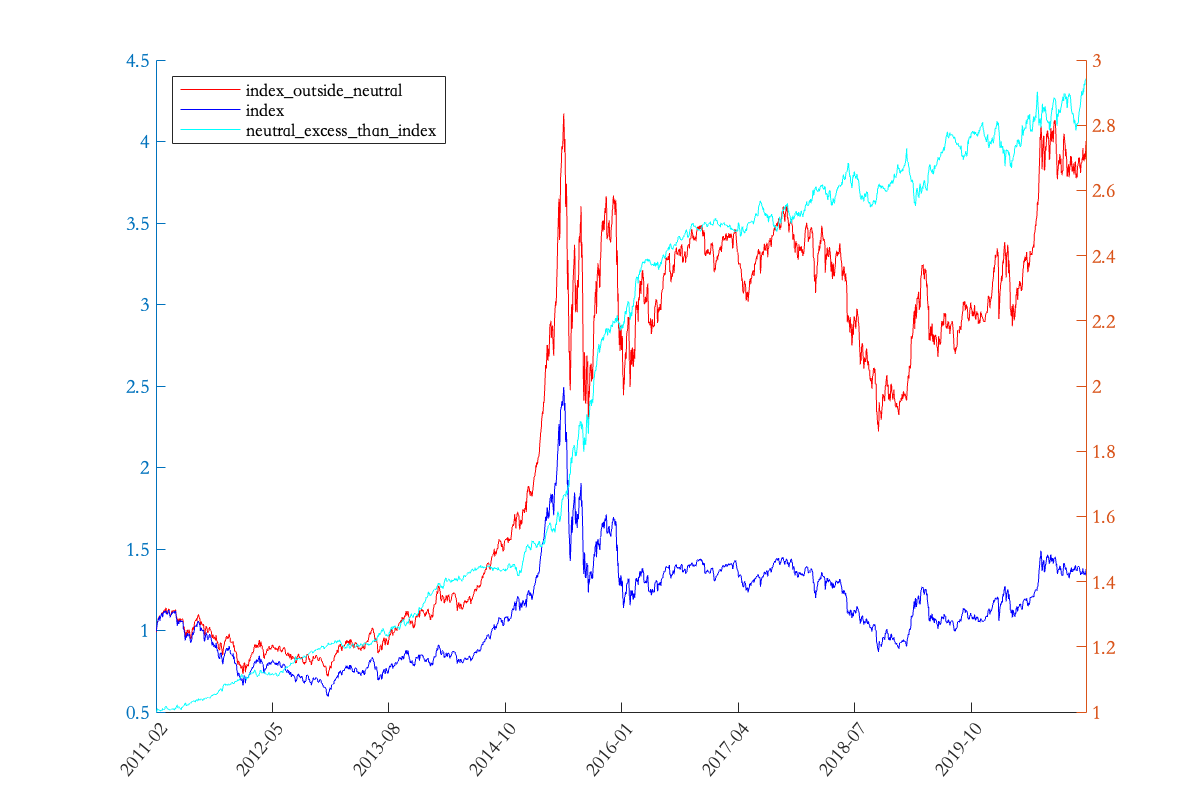}
    \caption{Performance of minimum absolute risk portfolio - stockmarket: all market}
    \label{fig:Fig3}
\end{figure}

Compared with CSI 500 index, our portfolio outperforms the index with lower annualized volatility, higher Sharpe ratio and smaller maximum drawdown. This indicates that the risk model we construct is relatively effective in portfolio optimization.

\subsubsection{Optimize difference of weight to minimize active risk}
In practice, passive index funds serve as a typical example of portfolio with minimum active risk. Here we want to construct portfolio with minimum active risk, and $w$ is the difference of weight compared to benchmark:
\begin{equation}
\begin{split}
    Min_w\ w^T(XFX^T+\Delta)w\\s.t.\ \forall n\ w_n+w^{bench}_n\ge0\\\Sigma_n(w_n+w^{bench}_n)=1
\end{split}
\end{equation}
\centerline{$w^{bench}_n$: benchmark index's portfolio weight of stock n}
~\\

Here we still choose CSI 500 index as benchmark. The selected portfolio from CSI 500 components is as below. Still, it outperforms the original index. Compared to the minimal absolute risk portfolio, minimal active risk portfolio's net value is closer to CSI 500 index, indicating that it has lower active risk than minimal absolute risk portfolio. This shows that the portfolios we obtain satisfied our initial optimization objective.

\begin{figure}[H]
    \centering
     \includegraphics[width=0.7\textwidth]{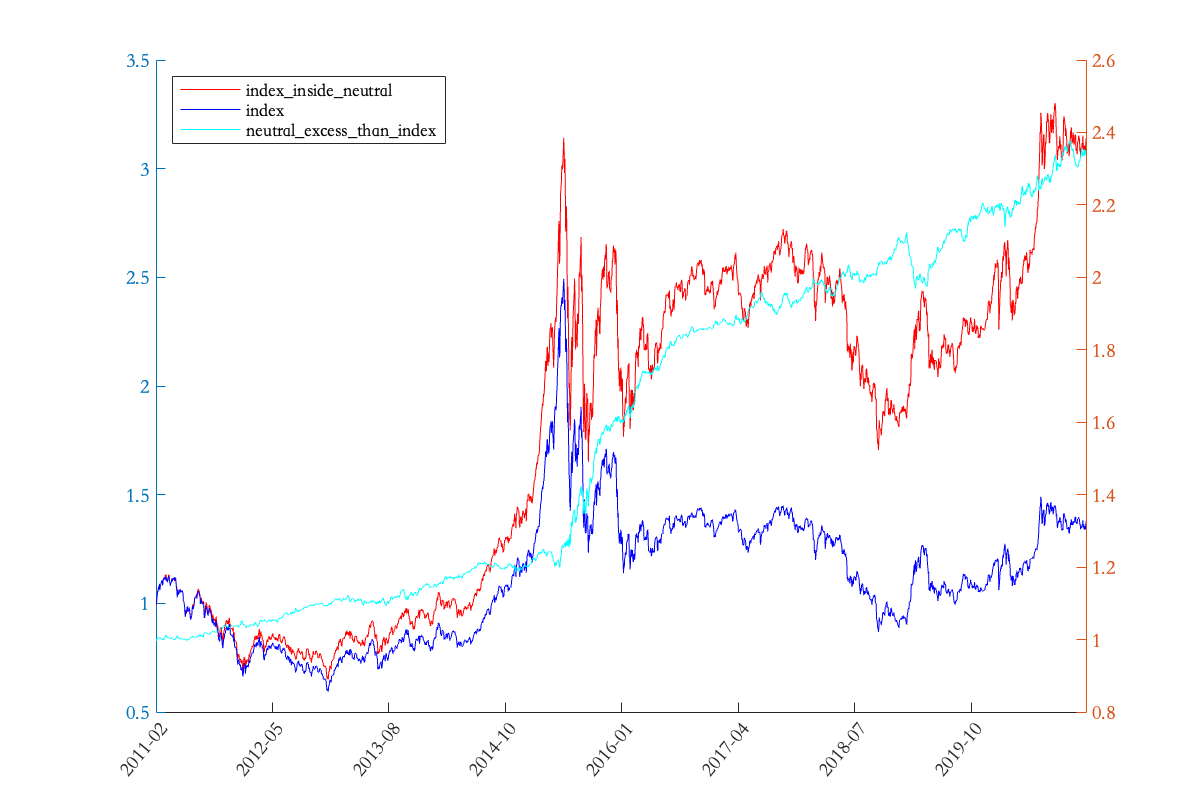}
    \caption{Performance of minimum active risk portfolio - stockmarket: index constituent stocks}
    \label{fig:Fig4}
\end{figure}

\subsection{Portfolio that maximizes risk-adjusted return}
Since investors are typically more concerned about return, it is more reasonable to construct portfolio with maximal return after adjusting for risk. The objective function for maximizing risk-adjusted return is as follows:
~\\
\begin{equation}
\begin{split}
    Max_w\ w^Tr-\lambda\cdot w^T(XFX^T+\Delta)w 
\end{split}
\end{equation}
\centerline{$\lambda$: risk aversion coefficient}
\centerline{$r$: expected return vector obtained from return model}
~\\
Still, if we want to maximize portfolio's risk-adjusted absolute return, $w$ is the portfolio weight. If we want to maximize portfolio's risk-adjusted active return, $w$ then represents the difference of weight compared with benchmark.

The amount of risk an investor is willing to take to obtain one unit of return is Risk aversion coefficient $\lambda$ is defined as the amount of an investor is willing to take in order to obtain one unit of return. Sharpe ratio can be expressed as follows:
~\\
\begin{equation}
\begin{split}
    SR=\frac{w^Tr}{\sigma_p(w)} =\frac{w^Tr}{\sqrt{w^T(XFX^T+\Delta)w}}
\end{split}
\end{equation}
\centerline{$w$: portfolio weight}
~\\
And the optimization objective can be expressed as:
\begin{equation}
\begin{split}
    Max_w\ SR\cdot \sigma_p(w)-\lambda\cdot\sigma^2_p(w)
\end{split}
\end{equation}
Objective function reaches its maximum value at $\sigma_p(w)^*=SR/(2\cdot\lambda)$, and the optimal value of risk aversion coefficient is:
~\\
\begin{equation}
\begin{split}
    \lambda^*=SR/[2\cdot\sigma_p(w)^*]
\end{split}
\end{equation}
~\\
Information ratio is written as:
~\\
\begin{equation}
\begin{split}
    IR=\frac{w^Tr}{\sigma_p(w)} =\frac{w^Tr}{\sqrt{w^T(XFX^T+\Delta)w}}
\end{split}
\end{equation}
\centerline{$w$: difference of weight compared with benchmark}
~\\
And the objective function can be expressed as:
\begin{equation}
\begin{split}
    Max_w\ IR\cdot \sigma_p(w)-\lambda\cdot\sigma^2_p(w)
\end{split}
\end{equation}
Still, objective function reaches its maximum value at $\sigma_p(w)^*=IR/(2\cdot\lambda)$, and the optimal value of risk aversion coefficient is:
~\\
\begin{equation}
\begin{split}
    \lambda^*=IR/[2\cdot\sigma_p(w)^*]
\end{split}
\end{equation}
~\\
Thus, we need to choose different risk aversion coefficients for different optimization problems, and we can obtain a rough range of risk aversion coefficient based on historical Sharpe ratio, information ratio and volatility of the stock pool.

Similar to the optimization of minimum risk portfolio, we can add extra constraints in our optimization to adjust for certain need or market regulations.

\subsubsection{Optimize portfolio weight to maximize absolute return}
In this subsection, $w$ is the portfolio weight and we want to construct a portfolio that maximizes risk-adjusted absolute return.
~\\
\begin{equation}
\begin{split}
    &Max_w\ w^Tr-\lambda\cdot w^T(XFX^T+\Delta)w\\
    &s.t.\ \forall n\ 0\leq w_n\leq0.01\\
    &\Sigma_nw_n=1\\
    &\forall i\ (w-w^{Bench})^TX_i=0\\
    &|(w-w^{Bench})^TX_{Size}|\leq0.5
\end{split}
\end{equation}
\centerline{$w^{Bench}$: weights of stocks in benchmark}
\centerline{$X_i$: all stocks' factor exposures on $i_{th}$ industry factor}
\centerline{$X_{Size}$: all stocks' factor exposures on $Size$ style factor}
~\\
Here the constraints are: short sell is not allowed, maximum weight for a single stock is $1\%$, the weight summation of all stocks is $100\%$, and the portfolio is neutral to industry and market cap of stocks. Benchmark is CSI 500 index and portfolio is constructed within its components. By setting the risk aversion coefficient to be 0, the performance of the optimized portfolio is as follows.

\begin{figure}[H]
    \centering
     \includegraphics[width=0.8\textwidth,height=0.4\textwidth]{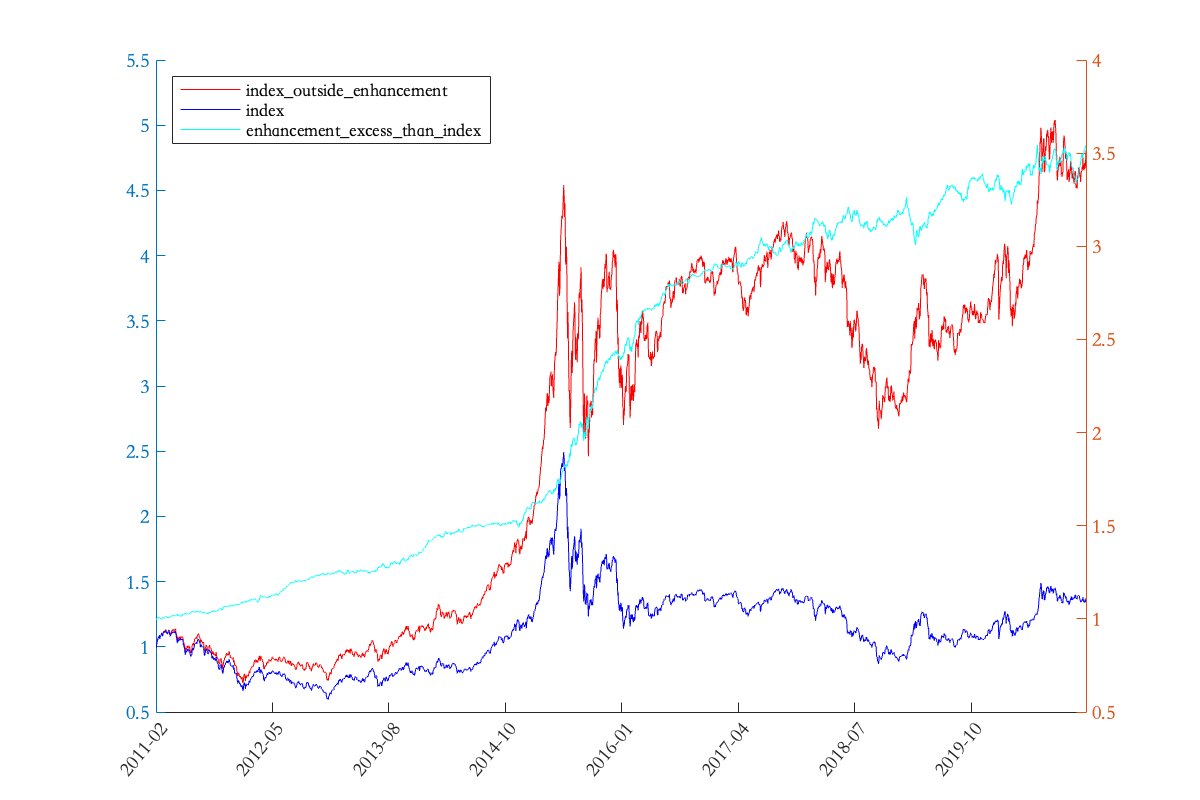}
    \caption{Performance of the portfolio maximizing risk-adjusted absolute return - stockmarket: all market}
    \label{fig:Fig12}
\end{figure}

\begin{figure}[H]
    \centering
     \includegraphics[width=0.8\textwidth,height=0.4\textwidth]{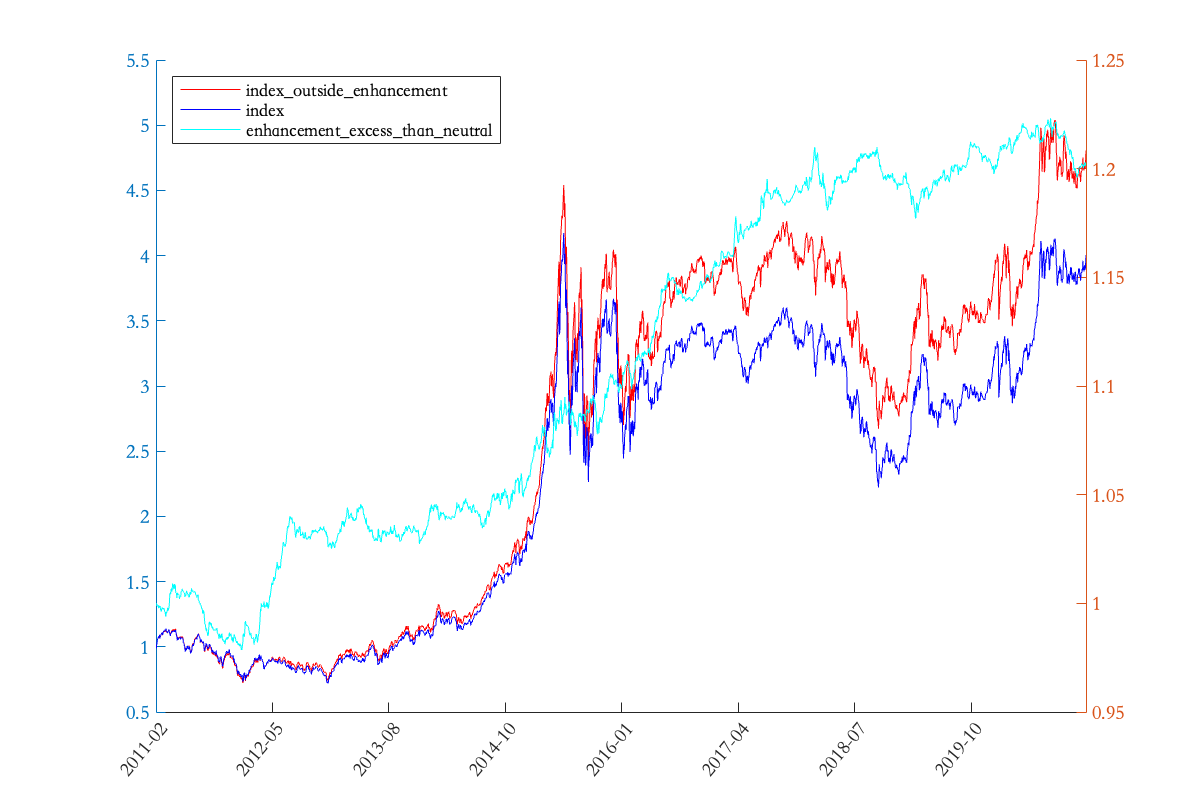}
    \caption{Comparison of portfolio that maximizes risk-adjusted absolute return and that minimizes absolute risk - stockmarket: all markets}
    \label{fig:Fig13}
\end{figure}

The portfolio that maximizes risk-adjusted absolute return outperforms the original index and the minimal absolute risk portfolio, and has higher return than both of them. However, it is more volatile. In other words, its higher return is partially due to risk premium.

\subsubsection{Optimize difference of weight to maximize active return}
In this subsection, $w$ is the difference of weight compared to benchmark. We want to construct a portfolio that maximizes risk-adjusted active return and discuss the effect of different risk aversion coefficient.


Setting the CSI 500 Index as benchmark, we build a portfolio selected within benchmark components. The constraint conditions include not allowed to short, $1\%$ as weight upper limit weight per stock, a $100\%$ weight total for all stocks, and the portfolio being neutral with respect to the benchmark and market cap. The optimal portfolio is written as 
~\\
\begin{equation}
\begin{split}
    &Max_w\ w^Tr-\lambda\cdot w^T(XFX^T+\Delta)w\\
    &s.t.\ \forall n\ \ 0\leq (w_n+w^{Bench})\leq0.01\\
    &\Sigma_n(w_n+w^{Bench})=1\\
    &\forall i\ w^TX_i=0 \\
    &|w^TX_{Size}|\leq0.5
\end{split}
\end{equation}
\centerline{$w^{Bench}$: weights of stocks in benchmark}
\centerline{$X_i$: all stocks' factor exposures on $i_{th}$ industry factor}
\centerline{$X_{Size}$: all stocks' factor exposures on $Size$ style factor}
~\\

Below is the performance of the portfolio that maximizes risk-adjusted active return. We can see that the the portfolio that maximizes risk-adjusted active return outperforms the original index and the minimum active risk portfolio. It has more volatile but higher net value than the CSI 500 index and the minimum active risk portfolio. This indicates that the maximum active return portfolio has higher return for taking more risk compared to the index and minimum risk portfolio.

\begin{figure}[H]
    \centering
     \includegraphics[width=0.8\textwidth]{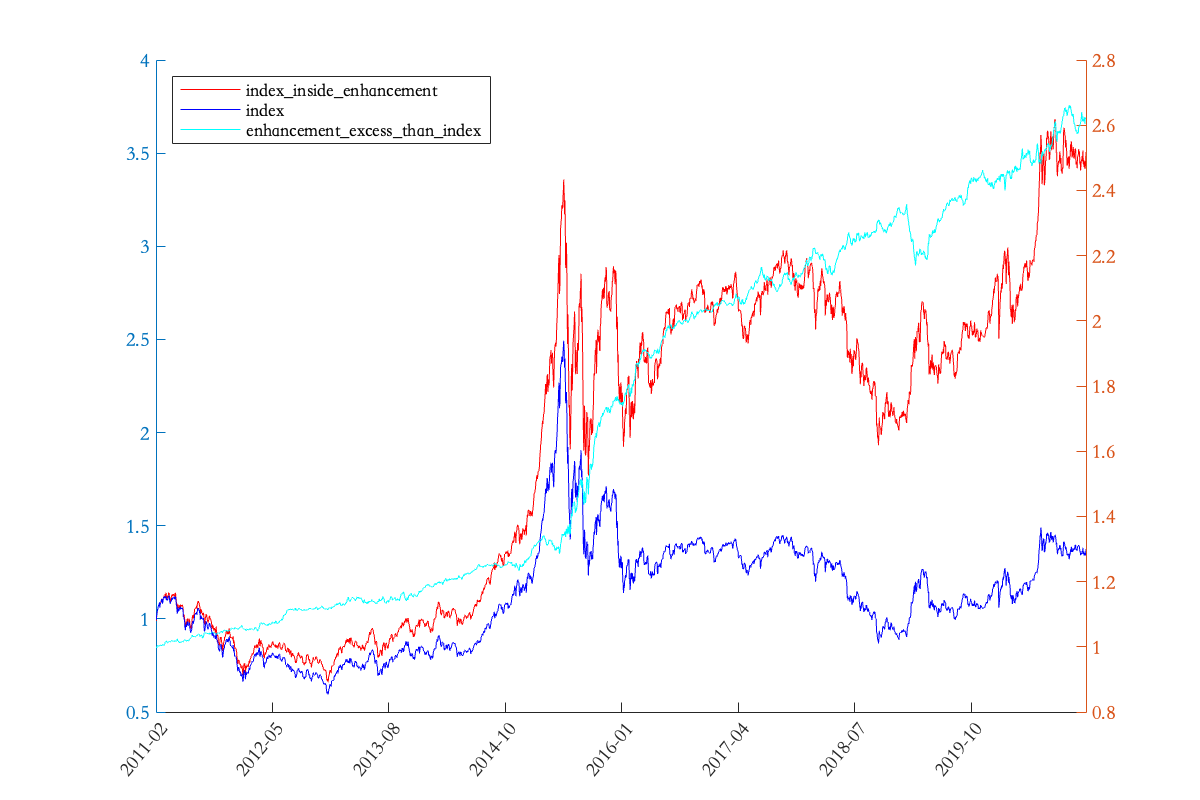}
    \caption{Performance of the portfolio maximizing risk-adjusted active return - stockmarket: index constituent stocks}
    \label{fig:Fig14}
\end{figure}

\begin{figure}[H]
    \centering
     \includegraphics[width=0.8\textwidth]{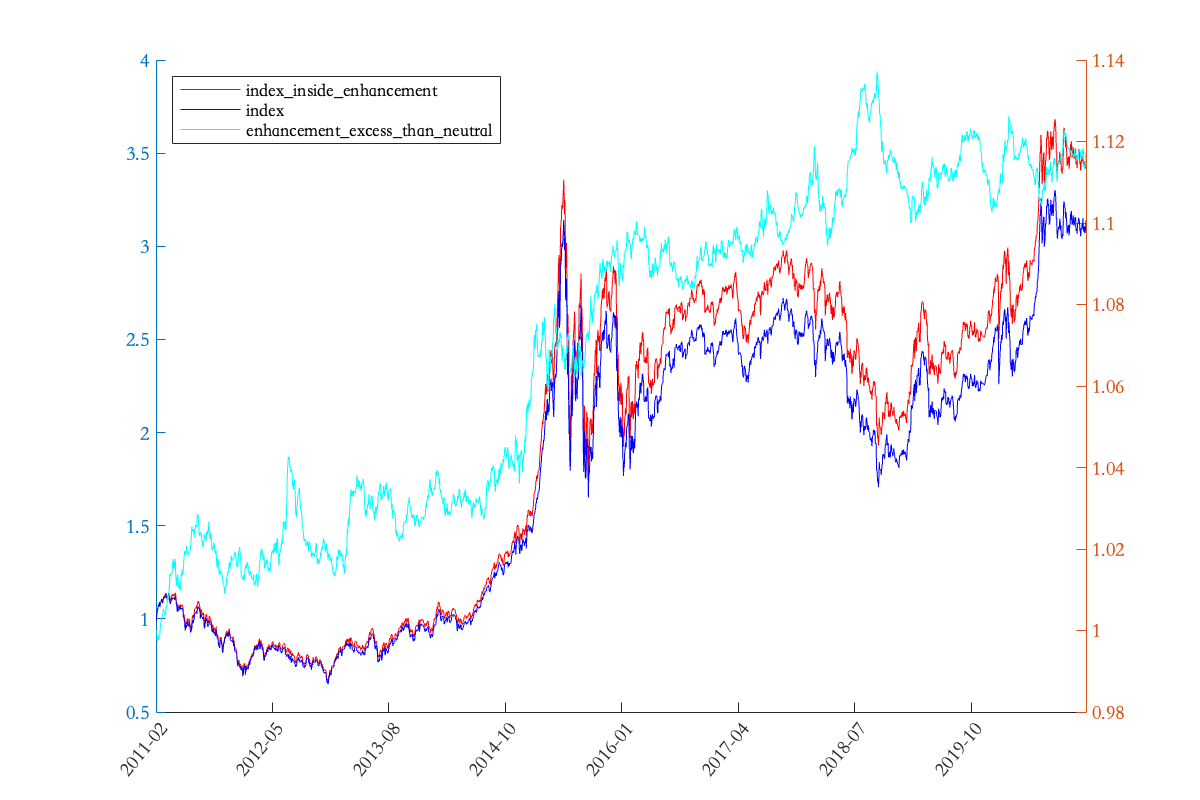}
    \caption{Comparison of portfolio that maximizes risk-adjusted active return and that minimizes active risk - stockmarket: index constituent stocks}
    \label{fig:Fig15}
\end{figure}

We evaluate the optimal portfolio with maximized active return for different risk aversion coefficients, and results are shown in the following charts. It is obvious that the portfolio perform better, concerning itself, its behavior compared with the least risk model and that compared with the benchmark, as $\lambda$ rises. To be more specific, the annualized return, Sharpe ratio and information ratio are higher while volatility and maximum drawdown become lower. What should be mentioned is that, despite a slight climb in volatility and maximum drawdown comparing with neutral model and the benchmark, the Sharpe ratio and information ratio always increase, which means that the optimal portfolio is well profitable. Also clearly, portfolio show better results when the success ratio increases.
\begin{figure}[htbp]
\centering
\subfigure[$\lambda=0$]{
    \includegraphics[width=17cm]{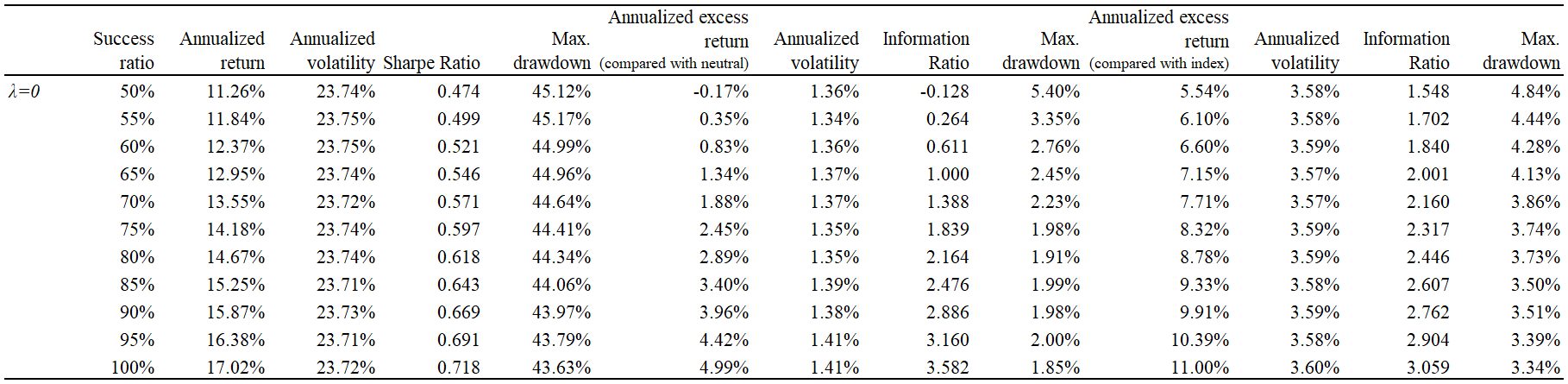}
}
\quad
\subfigure[$\lambda=0.5$]{
\includegraphics[width=17cm]{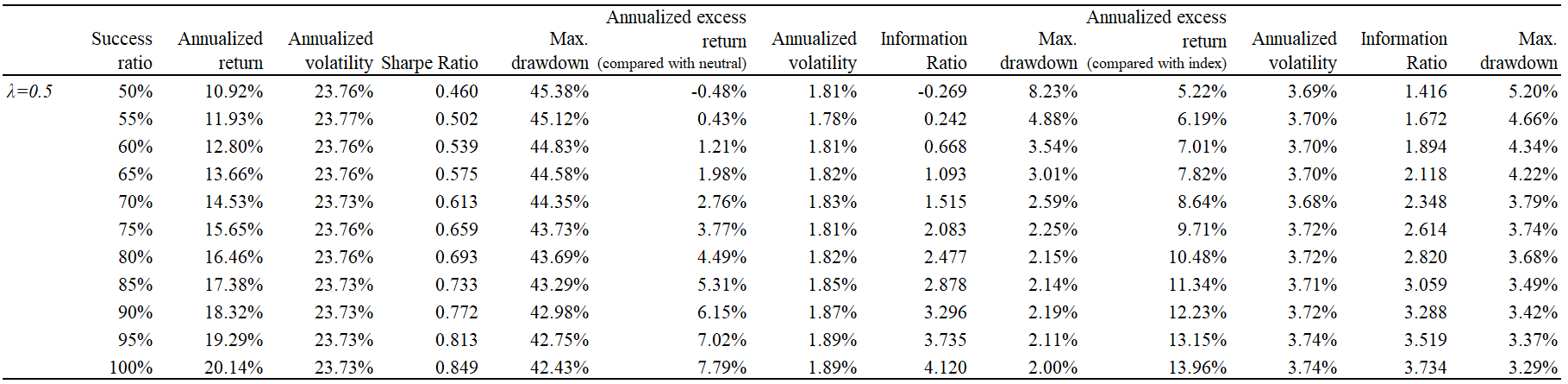}
}
\quad
\subfigure[$\lambda=1$]{
\includegraphics[width=17cm]{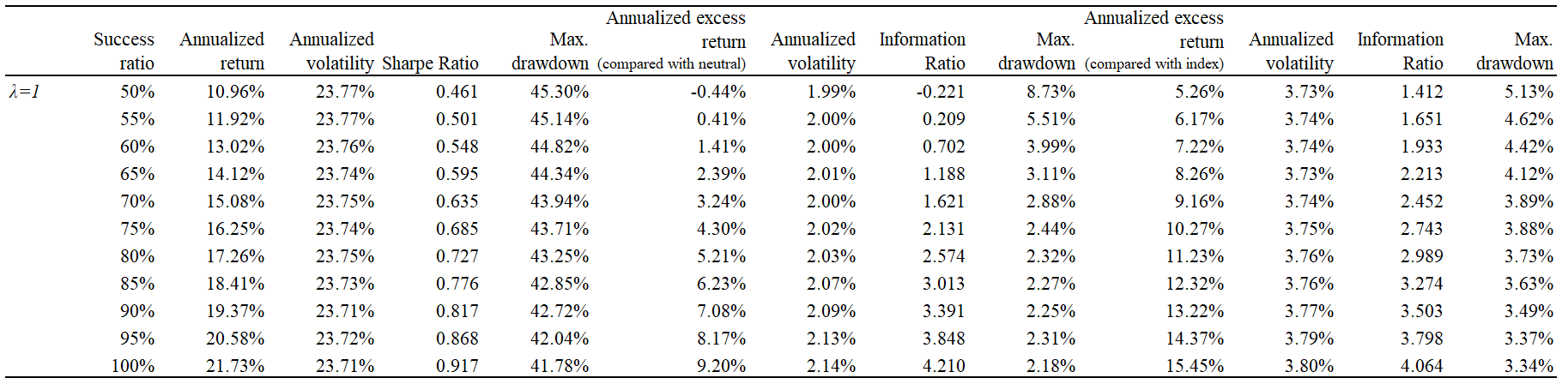}
}
\quad
\subfigure[$\lambda=2$]{
\includegraphics[width=17cm]{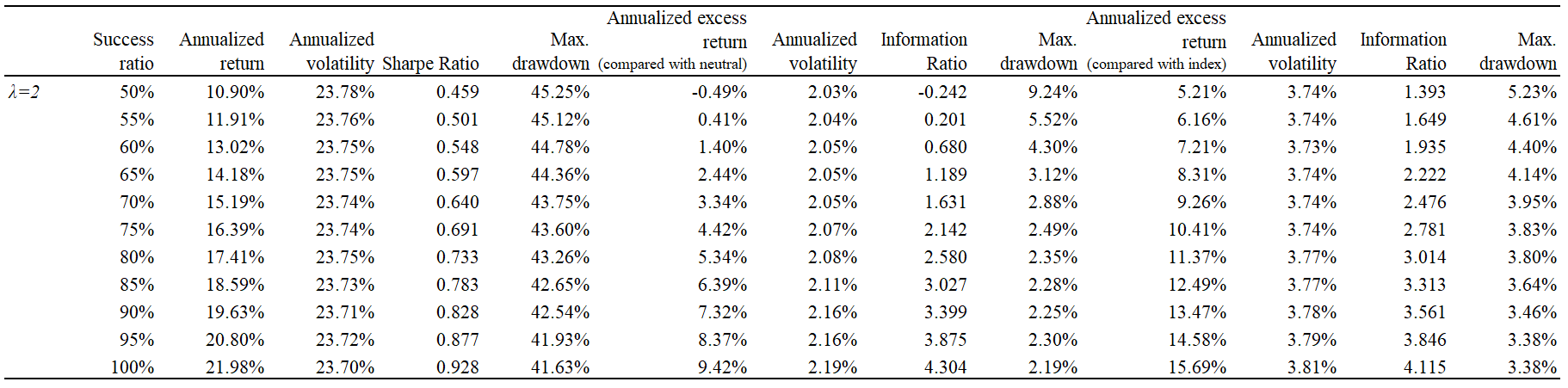}}
\caption{Performance indicators for optimal portfolio that maximize active return under different risk aversion coefficients}
\end{figure}
\\

\newpage
\section{Summary}

On the basis of the classical multi-factor model framework risk model, we propose a simple yet efficient Multi-factor risk model. We include 10 style factors, 29 industry factors and 1 country factor as risk factors, conduct cross sectional regression of the factor exposure and estimate the factor return covariance matrix and  idiosyncratic return variance matrix.

As for the estimation of the factor covariance matrix, we apply Newey-West adjustment to fix the error due to time-series autocorrelation of factor returns.
When estimating idiosyncratic return variance matrix, we introduce Newey-West adjustment and structural adjustment to correct the time series autocorrelation, variance matrix estimation error and errors from missing data and outliers. Adjusted models show good performance in risk prediction simulation with high accuracy.

Using the multi-factor risk model, We construct four portfolios that respectively aim to minimize absolute risk, active risk, maximize risk-adjusted absolute return and risk-adjusted active return.
Among them, the last portfolio, on the basis of records between February 1, 2011 and May 31, 2022, provides a at most 15.69\% excess return annually from benchmark CSI 500, with Sharpe ratio being 0.928, and the information ratio being 4.115. 
Compared with model without risk adjustment, the information ratio increased by more than 1 and Sharpe ratio by 0.21.

Our risk model is basically statistical risk model which lacks non-linear capacity and still deficient in out-of-sample prediction performance. So we still need to explore more effective and wide-adaptive ways to construct and adjust our risk model. In the future, we plan to explore more adjustment techniques, such as machine learning adjustments, to improve the estimation and help us select more effective risk factors.

\end{document}